# Quantitatively analyzing intrinsic plasmonic chirality by tracking the interplay of electric and magnetic dipole modes

Li Hu[a,c] Yingzhou Huang[c] and Yurui Fang[b*]

**Abstract:** Plasmonic chirality exhibits great potential for novel nanooptical devices due to the generation of a strong chiroptical response. Previous reports on plasmonic chirality explanations are mainly based on phase retardation and coupling. We propose a quantitative model similar to the chiral molecules for explaining the mechanism of the intrinsic plasmonic chirality quantitatively based on the interplay and mixing of electric and magnetic dipole modes, which forms a mixed electric and magnetic polarizability. The analysis method is also suitable for small chiral object down to quasi-static limit without phase delay and expected to be a universal rule.

Objects that cannot be superimposed with their mirror image are chiral. Chirality is a common concept in nature.[1,2] Circular dichroism (CD) is very sensitive to molecular conformations and chirality, which is the difference in absorption of left circularly polarized (LCP) light and right circularly polarized (RCP) light by chiral molecules. The CD spectroscopy is a powerful analytic tool ranging from the determination of sugar concentration in wine to quality control in pharmaceutically relevant processes. However, molecular CD of natural compounds such as organic and biological chiral molecules is typically very weak and occurs in the ultraviolet region (150-300 nm), which have limited greatly those applications. Proving CD effect of chiral molecules beyond the UV region with large magnitudes is thus of great interest. Recently, it has been suggested to use plasmonic nanostructures to boost the sensitivity of the method by generating super chiral electromagnetic near-fields or exploiting their acute response to their immediate environment.[3] A lot of works have already been done and demonstrated that the artificial plasmonic nanostructures can give rise to natural optical activity, such as chiral metal particle,[4,5] DNA based self-assembled metal particles,[6,7] helical nanowires,[8-10] etc.[11-13]

In the past several years, a lot of works have been proposed for explaining the chiral mechanism, such as the phase delay of two orthogonal directions of the structures,[14] the asymmetry of the structure to the circularly polarized light (CPL),[15] the resonant modes over lapping with the rotation direction of the CPL[16] and Born-Kuhn model,[17] etc. However, all of the explanations for the CD effect of the plasmonic structures are still phenomenological. A more general profound way is analogic to the plasmonic structure as a chiral molecule.[1] The chiral mechanism in the chiral molecule model is obtained from ab initio model that the chiral effect comes from the interaction of the electric and magnetic dipoles. Plum borrowed the idea and qualitatively explained the extrinsic chirality of plasmonic nanostructure.[18,19] Later, Tang borrowed Lipkin' theory[20] and developed the chirality of the chiral field. In our last work, we quantitatively analyse the mechanism extrinsic plasmonic chirality with the interplay of electric and magnetic dipole modes.[21] However, the analytical or quantitative explanation of the CD in intrinsic three-dimensional (3D) chiral nanostructure is still unavailable.

In this short communication, inspired by the chiral molecule theory, we present a similar analytical model and quantitatively analyse the plasmonic CD of 3D chiral nanostructure. The interplay of electric and magnetic modes of the nanostructure is analysed and considered to be responsible for the CD in dipole resonant range. The analytical model as well as the finite element method (FEM, COMSOL Multiphysics) analysis agrees with the CD quite well quantitatively. Compared the analytical model and the Born-Kuhn model, the results are consistent. In addition, some other 3D plasmonic chiral nanostructure also were used to investigate the analytical model. Higher order modes are out of consideration of this model, and some mismatch is possible caused by the electric dipole quadruple interaction.

As we all known, the CD of chiral molecule comes from the mixed interacting electric-magnetic dipole momentum and electric dipole-quadruple momentum.[1] In experiments, as plenty of

[a.] School of Computer Science and Information Engineering, Chongqing Technology and Business University, Chongqing, 400067, China.
[b.] Department of Physics, Chalmers University of Technology, Gorthenburg SE41296, Sweden. *The correspondence should be addressed to email: yurui.fang@chalmers.se
[c.] Soft Matter and Interdisciplinary Research Center, College of Physics, Chongqing University, Chongqing, 400044, P. R. China

Electronic Supplementary Information (ESI) available: Coupled dipoles approximation, numerical method, and extinction, CD and dipole components ($P_e$, $P_m$) spectra for different structures.

molecules are randomly distributed, the electric dipole-quadruple interaction is cancelled by average. Analogic to the chiral molecules, for the 3D chiral plasmonic structure, the CD effect results from the interaction of both electric (e)- magnetic (m) dipoles and electric dipole-quadruple (q) responses of the structure.[1] For higher order modes, they usually can be decomposed into the three basic modes in different parts of the structure in simple chiral plasmonic structures. In most of the chiral structure, the electric quadruple mode is equivalent to the magnetic mode. So we mainly focus on the interplay of the electric and magnetic modes. When the electric dipole and magnetic dipole can interact with each other, where the electric dipole, magnetic dipole and wave vector construct a three-dimensional chiral configuration, there will be CD effect. For the extrinsic chirality, the interaction occurs with the assistant of the titled incident electromagnetic (EM) wave.[19, 21] While for the 3D intrinsic chirality, the electric and magnetic dipoles have a non-orthogonal angle, the dipoles have non-zero projection on each other, which causes a non-zero dot product. When the electric and magnetic dipole modes in the chiral structure are excited simultaneously by the incident EM field, there is a mixed electric-magnetic dipole polarizability $\mathbf{G} = \mathbf{G}' + i\mathbf{G}''$, which makes the electric dipole moment $\mathbf{p}_e$ and magnetic dipole moment $\mathbf{p}_m$ as

$$\widetilde{\mathbf{p}_e} = \tilde{\alpha}\widetilde{\mathbf{E}} - i\tilde{G}\widetilde{\mathbf{B}}, \qquad \widetilde{\mathbf{p}_m} = \tilde{\chi}\widetilde{\mathbf{B}} + i\tilde{G}\widetilde{\mathbf{E}},$$

Where $\tilde{\alpha} = \alpha' + i\alpha''$ is the electric polarizability and $\tilde{\chi} = \chi' + i\chi''$ is the magnetic susceptibility. $\mathbf{E}$ and $\mathbf{B}$ are the local fields at the meta-molecule.

The extinction of the nanostructure is given by[21, 2]

$$\sigma^\pm = \frac{\omega}{2} Im(\mathbf{E}^* \cdot \mathbf{p}_e + \mathbf{B}^* \cdot \mathbf{p}_m) = \frac{\omega}{2}\left(\alpha''|\widetilde{\mathbf{E}}|^2 + \chi''|\widetilde{\mathbf{B}}|^2\right) + G''^\pm \omega\, Im(\widetilde{\mathbf{E}}^{\pm *} \cdot \widetilde{\mathbf{B}}^\pm) \quad (1)$$

So

$$\Delta\sigma = G''^+ C^+ - G''^- C^- \quad (2)$$

where $C = -\frac{\varepsilon_0 \omega}{2} Im(\mathbf{E}^* \cdot \mathbf{B})$ is the optical chirality, which can be calculated for any monochromatic electromagnetic field. $C_{CPL} = \pm \frac{\varepsilon_0 \omega}{2c} E_0^2$ is the optical chirality for a right (−) or left (+) circularly polarized plane wave.

The main point here is that the $\mathbf{G}$ is usually unknown (especially in the complex structures, in a lot of related papers a chiral parameter κ is directly assumed to be a fixed value), and we find that a direct dot product of electric and magnetic dipole modes of the chiral plasmonic structures can be used to express the $\mathbf{G}$ value just as a molecule. Different from the extrinsic chirality reported last time with $G'' \sim -Im(\mathbf{p}_e^* \cdot (-i\mathbf{p}_m))$,[21-23] we find that for the intrinsic chirality we have

$$G'' \sim -Im(\mathbf{p}_e^* \cdot \mathbf{p}_m) \quad (3)$$

because of the simultaneous excitation of electric and magnetic dipoles by in plane $\mathbf{E}$ and $\mathbf{H}$ components of the EM wave. For the extrinsic chirality, as the wave vector, induced electric mode and magnetic dipole mode ($\mathbf{k, d, m}$) are in the same plane, so the interaction of $\mathbf{m}$ and $\mathbf{H}$ has a π difference with the $\mathbf{d}$ and $\mathbf{E}$ interaction. However, for intrinsic chirality, $\mathbf{d}$ and $\mathbf{m}$ can directly interact in the plane perpendicular with the wave vector $\mathbf{k}$, so they can interact with $\mathbf{E}$ and $\mathbf{H}$ simultaneously, so *there is a i factor difference here for the plasmoic intrinsic and extrinsic chirality*.

We first analyse the intrinsic plasmonic chiral molecule conception with a coupled-dipole approximation model as shown in Figure 1a with the formula we use before (see Supporting information).[21] The incident light propagating against z axis. The polarizability of the electric dipole $\overrightarrow{\alpha_1}$ is set as a gold ellipsoid particle (a = 100 nm, b = c = 30 nm) with long axis along $\mathbf{x}$. Here $\overrightarrow{\mathbf{u}_2}$ is set on the basis of an ellipsoid particle[‡] (a = 30 nm, b = c = 4.5 nm. ) with long axis along an azimuth angle by 20° to z axis and 45° to x axis  (Figure 1a). The uncoupled e and m dipoles momenta are shown in Figure 1b and *they are pure e or m dipoles without the mixed polarizability $\mathbf{G}$*. When they are coupled together, the dipoles momenta of the coupled system (Figure 1c) clearly show *the mixed polarizability*. Both the resonant peaks have electric and magnetic moments mixed. So there are interactions of electric and magnetic components in both electric and magnetic hybrid modes under incident EM wave. The coupled system shows clear CD effect as expected because of the mixed polarizability of the electric and magnetic dipoles, just like the chiral molecules. The extinction under CPL is got from $\sigma^\pm = \frac{\omega}{2} Im(\mathbf{E}^* \cdot \mathbf{p}_e + \mathbf{B}^* \cdot \mathbf{p}_m)$ as shown in Figure 1b. The two peaks split and show a Fano profile with a dip right at the individual magnetic resonant wavelength because the energy transfer between the e and m dipoles with phase delay, which is a typical Fano interference process. *As both CD and Fano can be caused by the interaction of e and m modes, usually when there is CD effect, there is possibly Fano effect.* The CD is a direct difference of the extinction shown in Figure 1d blue curve. On the other hand, the CD calculated with the molecule analysis formula 2 and (3) is shown in Figure 1d red curve. The two curves match each other very well. This indicates that the CD generation mechanism of the intrinsic plasmonic 3D chiral structures is quantitatively determined and connected to the CD spectrum by the interaction of mixed of e and m dipole modes.

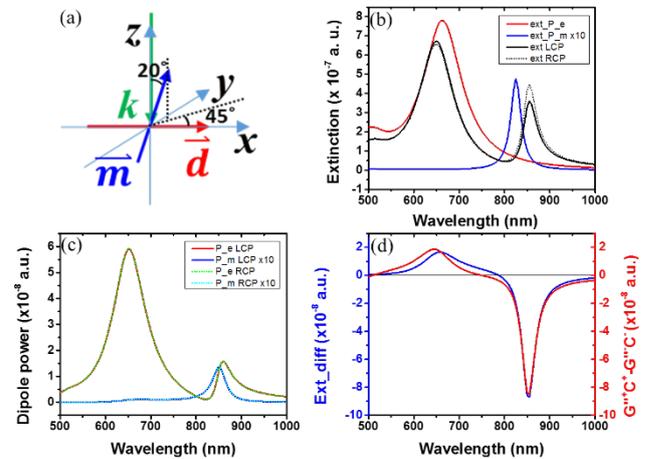

**Figure 1**. Coupled dipole approximation calculations for coupled electric dipole (ellipsoid a = 100 nm, b = c = 30 nm) and magnetic plasmonic dipole (ellipsoid a = 30 nm, b = c = 4.5 nm). (a) Schematics of the orientation of the wave vector, the coupled electric dipole and magnetic dipole. (b) Extinction spectra for uncoupled electric and magnetic plasmonic dipoles, as well as coupled electric and magnetic plasmonic dipoles under CPL illuminations. (c) The dipole power of the individual coupled electric and magnetic plasmonic dipoles. (d) Extinction difference (CD) of the coupled system (blue curve) and imaginary part of the mixed electric and magnetic polarizability of the coupled system

One of the most intuitive way to classically understand the generation of natural optical activity in chiral media is provided by the coupled oscillator model of Born and Kuhn[17, 24, 25] which is consisting of two identical, vertically displaced, coupled oscillators, as shown in Figure 2a inset. The plasmonic Born-Kuhn model is investigated with the same way as the mixed coupled dipoles analytical model but with electric and magnetic dipoles momenta obtained from numerical FEM results (Figure 2). Figure 2a shows the extinction spectrum of the structure under LCP and RCP excited. The electric and magnetic dipole moment are obtained with

$$\boldsymbol{p}_e = \int d^3 r' \boldsymbol{r}' \rho(\boldsymbol{r}') \quad (4)$$

$$\boldsymbol{p}_m = \frac{1}{2} \int d^3 r' \, (\boldsymbol{r}' \times \boldsymbol{J}) \quad (5)$$

where $\rho(\boldsymbol{r}')$ is the charge density and $\boldsymbol{J}(\boldsymbol{r}')$ is the current density. Figure 2b shows the derived electric and magnetic dipole moments. Like previous proved, the electric and magnetic dipole resonances are mixed together, showing electric and magnetic components at the resonant peaks. Then the CD is got as traditional way for extinction difference, and the mixed polarizability is obtained with Formula 2 and 3 (we only show G'' here, as the chiral fields $C_{CPL}$ for LCP and RCP have opposite sign, so G'' is the sum of $G''^{\pm}$)[†]. From Figure 2c, we can see that the CD spectrum which is calculated with Formula 3 (red curve) and the extinction difference (blue curve) are consistent. The result confirms our deduction quantitatively analytical mechanism, which is a further step in explaining the plasmonic 3D chiral structures compared with the previous one. The bisignate usually can be easily explained by the peak shift under LCP and RCP excitations. In the coupled e and m model, it is can also be explained as the projected magnetic dipole component on electric dipole (dot product, see Supporting Information Figure S1) is in the same direction or opposite direction. However, this approach only works for dipole modes.

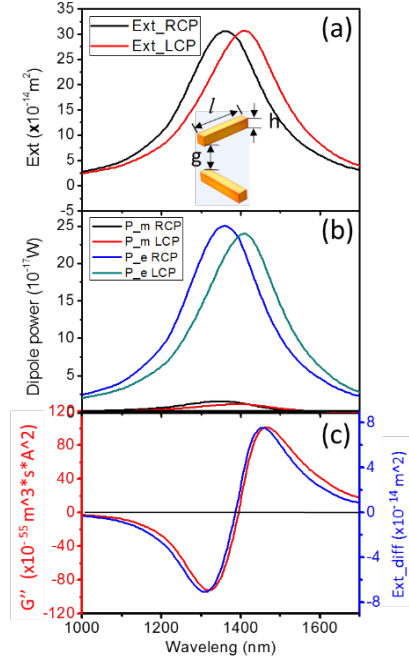

**Figure 2**. Coupled electric and magnetic dipoles analysis for the Born-Kuhn model. (a) Extinction spectra of the structure (inset) under LCP and RCP excited (length $l$ = 223 nm, height $h$ = 40 nm, width $w$ = 40 nm, gap $g$ = 120 nm). (b) The electric and magnetic dipoles power yielded by the structure under CPL illumination. (c) The imaginary part of the mixed electric and magnetic polarizability (red curve) and the extinction different (CD) of the coupled system (blue curve).

To further verify the above mechanism, six plasmonic chiral structures which have been published in other papers are investigated with the analytical model (see Supporting Information for details). The electric and magnetic dipoles momenta are obtained from FEM numerical results with formula 4 and 5. And the CD spectra from formula 3 (black curve) and from extinction difference (blue curve) are compared in Figure 3. First, the quasi-three-dimensional chiral heterotetramers is investigated with analytical model (Figure 3a). The CD effect is due to strong near field coupling and intricate phase retardation effects qualitatively as said in Ref[26]. From Figure 3a we can see that the dipole modes match well but not so for multiple modes, because for multiple modes there are more than one equivalent magnetic dipoles oscillating out of phase, but with the delayed interaction with the electric modes, thus the whole effect is interference superposition. Figure 3b shows the CD of plamonic nanohelix which had been analyzed by discrete dipole approximation method and explained as the extinction combination of under x and y polarization with complementary phases.[14, 27] The CD spectra from formula 3 and extinction difference anastomosed each other. Figure 3c shows the three-dimensional arrangements of plasmonic "meta-atoms" which is because of the chiral configuration caused helical displacement currents giving rise to an induced magnetic moment component parallel to the usual electric dipole moment and resulting an unique interaction of the meta-molecule with CPL.[28] The two curve match well. Figure 3d shows the compare for a spherical nanoparticle with a slight deformations. The CD mechanism has been explained with the mixing between plasmon harmonics with different angular momenta.[5] Here the two curve match the main trend but with larger deviation possibly because there are other non-dipole modes appearing significant. Figure 3e shows a 3-D chiral nanocrescent and the CD comes from the rotation incident E field vector matches the arranged equivalent dipoles at the resonant modes.[16] The CD curve from the coupled dipole modes analysis match the extinction result very well. For the self-assembled gold nanohelix as shown in Figure 3f, it shows CD because the helical arrangement of the gold nanoparticles result in coupled plasmon waves propagating along a helical path and causing increased absorption of those components of the incident light that are in accord with the handedness of the helices.[29] We can see that the spectrum of G'' and extinction difference are coincident for dipole modes. The slight mismatch may come from the quadruple modes arranged by the small particle dipole which cannot be totally treated as magnetic mode. From the above analysis, it can be seen that the CD of the 3D plasmonic structures can be quantitatively presented by the coupled e and m dipoles analytical model. For a few dipole models there are little difference because of the interaction of dipole and quadruple mode, where the quadruple mode cannot be totally represented by a magnetic dipole.

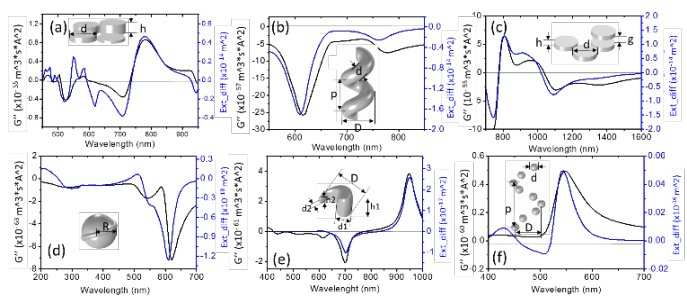

**Figure 3**. The comparison of the CD spectra for the coupled electric-magnetic dipole modes analysis and the extinction difference of 3D plasmonic chiral structures. (a) Quasi-three-dimensional oligomers (d = 100 nm, h = 10 nm, 20 nm, 30 nm, 40 nm, the gap between the oligomers: 2 nm). (b) Plasmonic nanohelix (major diameter: D = 36 nm, minor diameter: d = 28 nm, helical pitch: p = 60 nm).(c) 3D chiral plasmonic oligomers (diameter: d=100 nm, thickness of nanoparticle: h = 40 nm, the gap between the oligomers in borrom layer: s = 20 nm, the gap of two layers: g = 70nm). (d) Chiral nanocrystals (radius of sphere: R = 7 nm, radius of the twiser: 1 nm). (e) Spiral-type ramp nanostructures (outer diameter D = 22.5 nm, root diameter d1 = 11.3 nm, tip diameter d2 = 1.875 nm, the height at the root h1 = 17.5 nm, the height at the tip h2 = 2.5 nm). (f) Gold nanoparticle helices nanosphere (diameter: d =10 nm, major diameter: D = 34 nm, helical pitch: p = 54 nm).

In summary, the strong interplay of the electric and magnetic modes result in the CD of plasmonic 3-D intrinsic chiral structures and the modes represented e and m dipoles can be use quantitatively analyze the CD mechanism with the imagery part of mixed e and m polarizability G, which is a dot product of the two dipoles. The mixed polarizability comes from the strong interplay of electric and magnetic dipoles. The model is presented with coupled dipole approximation and verified with numerical FEM result in Born-Kuhn model (which is a classic model for chirality) and other 3D plasmonic chiral nanostructures. It is not suitable for higher order modes. Mismatch appears for some structures because the interaction of dipole and quadrupole modes are not ignorable, which is a continuous issue in the future. The quantitative model is expected to be applied to all the 3D chiral structures. The ways of getting mixed polarizability G for intrinsic and extrinsic plasmonic chirality[21] have a phase difference may be because the magnetic mode is excited by the H field with different phase, which also needs to be figured out in the future.

## Notes and references

‡ Generally, the magnetic response of most of the materials in the visible light range cannot catch up the fast altering light oscillation, so there is not proper material that has negative magnetic susceptibility and resonance at this range. Thanks for plasmonics, the oscillating equivalent circular current in the plasmonic structure can be treated as magnetons, and they have as fast response as the visible light frequency and equivalent negative magnetic susceptibility, known as chiral magnets.[30] So with certain arranged plasmonic structure, the induced dark mode is like a strong resonant magneton, where the magnetic susceptibility of the equivalent magnetic dipole $\overleftrightarrow{u}$ can be treated like plasmonic ellipsoid particle.

† $G$ is the mixed electric and magnetic polarizability, whose unit is $m^3/\Omega$ (which corresponding to $\alpha_{em}[m^3]/Z_0[\Omega]$), and Consider the expression of $C = -\varepsilon_0\omega/2 * Im(\boldsymbol{E}^* \cdot \boldsymbol{B})$, $G \to G/\varepsilon_0$, which has the unit $\alpha_{em}[m^3] * c[m/s]$. While in the calculation we only used $\boldsymbol{p_e} \cdot \boldsymbol{p_m}$ whose units is $m^3sA^2$, the chiral response for molecule is basically $\kappa = 4\pi n_0/3 * (2/\hbar \sum [\omega/(\omega_{jn}^2 - \omega^2) * (\langle n|\boldsymbol{d}|j\rangle\langle j|\boldsymbol{m}|n\rangle)])/\varepsilon_0$ (where $n_0$ is the volume density and $(2/\hbar \sum [\omega/(\omega_{jn}^2 - \omega^2)]$ is the density of states (DOS) ) which indicates that $G = (\boldsymbol{p_e^*} \cdot \boldsymbol{p_m}) * DOS = 2/(\hbar Z_0 c^2)\omega \, \boldsymbol{p_e^*} \cdot \boldsymbol{p_m} * [\boldsymbol{n_u} \cdot Im(\overleftrightarrow{G_e}(\boldsymbol{r_0},\boldsymbol{r_0};\omega_0) \cdot \boldsymbol{n_m})]$, where $\overleftrightarrow{G_e}$ has the relationship with electric field normal modes $\overleftrightarrow{G_e}(\boldsymbol{r},\boldsymbol{r'};\omega) = \sum_k c^2 \boldsymbol{e_k^*}(\boldsymbol{r'},\omega_k)\boldsymbol{e_k}(\boldsymbol{r},\omega_k)/(\omega_k^2 - \omega^2)$.[31] For simplifying in the simulations with FEM, we only calculated $\boldsymbol{p_e} \cdot \boldsymbol{p_m}$, but it doesn't affect the understanding for the mechanism.

# Quantitatively analyzing intrinsic plasmonic chirality by tracking the interplay of electric and magnetic dipole modes

Li Hu[a,c] Yingzhou Huang[c] and YuruiFang[b]

[a]School of Computer Science and Information Engineering, Chongqing Technology and Business University, Chongqing, 400067, China.
[b]Department of Physics, Chalmers University of Technology, Gorthenburg SE41296, Sweden. The correspondence should be addressed to email: yurui.fang@chalmers.se
[c]Soft Matter and Interdisciplinary Research Center, College of Physics, Chongqing University, Chongqing, 400044, P. R. China

**The electric and magnetic dyadic Green's functions**

The electric dyadic Green's tensor $\overleftrightarrow{G_e}(r,r_0)$ and magnetic dyadic Green's function tensor $\overleftrightarrow{G_m}(r,r_0)$ are the free space field susceptibility tensors propagator relating an electric dipole source $p_e$ at position $r_0$ in vacuum to the electric field $E$ and magnetic field $H$ it generates at position $r$ through

$$E(r) = \frac{k^2}{\varepsilon_0} \overleftrightarrow{G_e}(r,r_0) p_e \quad (S1)$$

$$H(r) = ck^2 \overleftrightarrow{G_m}(r,r_0) p_e \quad (S2)$$

For the electric and magnetic fields generated by a magnetic dipole $p_m$

$$E(r) = -Z_0 k^2 \overleftrightarrow{G_m}(r,r_0) p_m \quad (S3)$$

$$H(r) = k^2 \overleftrightarrow{G_e}(r,r_0) p_m \quad (S4)$$

With

$$\overleftrightarrow{G_e}(r,r_0) = \frac{e^{ikr}}{r}\left[(\hat{n}\otimes\hat{n}-\overleftrightarrow{I}) + \frac{ikr-1}{k^2 r^2}(3\cdot\hat{n}\otimes\hat{n}-\overleftrightarrow{I})\right] \quad (S5)$$

$$\overleftrightarrow{G_m}(r,r_0) = \frac{e^{ikr}}{r}\left(1+\frac{i}{kr}\right)(\hat{n}\times\overleftrightarrow{I}) \quad (S6)$$

$$\overleftrightarrow{G_e}(r,r_0)p = \frac{1}{4\pi}\frac{e^{ikr}}{r}\left[(\hat{n}\times p)\times\hat{n} + \frac{ikr-1}{k^2 r^2}(3\cdot\hat{n}(\hat{n}\cdot p)-p)\right] \quad (S7)$$

$$\overleftrightarrow{G_m}(r,r_0)p = \frac{e^{ikr}}{r}\left(1+\frac{i}{kr}\right)(\hat{n}\times p) \quad (S8)$$

Where $r = |r-r_0|$, $k = 2\pi/\lambda$, $\hat{n} = \frac{r-r_0}{r}$.

**The coupled dipole approximation method**

Let us consider many three dimensional dipole scatters. The local field at each dipole can be expressed as[1]

$$p_{e,j} = \varepsilon_0 \overleftrightarrow{\alpha_j} E_{j,total} = \varepsilon_0 \overleftrightarrow{\alpha_j}\left(E_{j,in} + \sum_{k=1,k\neq j}^{N}\left(\frac{k^2}{\varepsilon_0}\overleftrightarrow{G_e}(r_j,r_k)p_{e,k} - Z_0 k^2 \overleftrightarrow{G_m}(r,r_0)p_{m,k}\right)\right) \quad (S9)$$

$$p_{m,j} = \overrightarrow{u_j} H_{j,total} = \overrightarrow{u_j}(H_{j,in} + \sum_{k=1,k\neq j}^{N}(ck^2 \overrightarrow{G_m}(r_j,r_k)p_{e,k} + k^2 \overrightarrow{G_m}(r,r_0)p_{m,k})) \quad (S10)$$

For coupled one electric and one magnetic dipoles, we have

$$p_e = \varepsilon_0 \overrightarrow{\alpha_1}(E_{1,in} - Z_0 k^2 \overrightarrow{G_m}(r_e, r_m)p_m) \quad (S11)$$

$$p_m = \overrightarrow{u_2}(H_{2,in} + ck^2 \overrightarrow{G_m}(r_m, r_e)p_e) \quad (S12)$$

We can easily get the self-consistent form of dipole moments

$$p_e = \frac{\varepsilon_0 \overrightarrow{\alpha_1} E_{1,in} - \varepsilon_0 Z_0 k^2 \overrightarrow{\alpha_1}\overrightarrow{G_m}(r_m,r_e)\overrightarrow{u_2}H_{2,in}}{\overrightarrow{I} + \varepsilon_0 c Z_0 k^4 \overrightarrow{\alpha_1}\overrightarrow{G_m}(r_m,r_e)\overrightarrow{u_2}\overrightarrow{G_m}(r_e,r_m)} \quad (S13)$$

$$p_m = \frac{\overrightarrow{u_2}H_{2,in} + \varepsilon_0 c k^2 \overrightarrow{u_2}\overrightarrow{G_m}(r_e,r_m)\overrightarrow{\alpha_1}E_{1,in}}{\overrightarrow{I} + \varepsilon_0 c Z_0 k^4 \overrightarrow{u_2}\overrightarrow{G_m}(r_e,r_m)\overrightarrow{\alpha_1}\overrightarrow{G_m}(r_m,r_e)} \quad (S14)$$

The extinction is

$$A^{\pm} = \frac{\omega}{2} Im(E^* \cdot p_e + B^* \cdot p_m) \quad (S15)$$

Radiation power of the dipoles

The radiation power expressions of the electric dipole $p_e$ and magnetic dipole $p_m$ are

$$Q_e = \frac{\omega^4}{12\pi\varepsilon_0 c^3}|p_e|^2 \quad (S16)$$

$$Q_m = \frac{\omega^4 Z_0}{12\pi c^4}|p_m|^2 \quad (S17)$$

**The polarizability of the ellipsoid dipole**

For an ellipsoid the polarizability tensor is

$$\overrightarrow{\alpha}(r,\omega) = \overrightarrow{\alpha_0}(r,\omega)[\overrightarrow{I} - \left(\frac{2}{3}\right)ik_0^3 \overrightarrow{\alpha_0}(r,\omega) - k^2/\overrightarrow{\alpha_0}]^{-1} \quad (S18)$$

where $\overrightarrow{\alpha_0}(r,\omega)$ is the Clausius-Mossotti polarizability

$$\overrightarrow{\alpha_0} = \begin{pmatrix} \alpha_1 & 0 & 0 \\ 0 & \alpha_2 & 0 \\ 0 & 0 & \alpha_3 \end{pmatrix} \quad (S19)$$

$$\alpha_j = 4\pi abc \frac{(\varepsilon_{particle} - \varepsilon_{medium})}{(3\varepsilon_{particle} + 3L_j(\varepsilon_{particle} - \varepsilon_{medium}))} \quad (S20)$$

$$L_j = \frac{abc}{2}\int_0^{\infty} \frac{dq}{(x^2+q)f(q)} \quad (S21)$$

with $j = a, b, c$, $f(q) = [(q+a^2)(q+b^2)(q+c^2)]^{1/2}$ and $a, b, c$ are the axis of the ellipsoid[2].

$\overrightarrow{u}$ is obtained in the same way. Because nature material has bad magnetic response in optical frequency and the magnetons in our paper are yield by the plasmon resonance with circular current, we used a fake $u_{particle}$ value to yield the magnetic resonance in optical wavelength range, which is from $\varepsilon_{Au}$ but with imaginary part divided by 1.5 (the magnetic mode is dark, so the spectrum profile is narrower).

**The mixed electric and magnetic dipole polarizability**

Comparing S13, S14 with $\widetilde{p_e} = \widetilde{\alpha}\widetilde{E} - i\widetilde{G}\widetilde{B}$, $\widetilde{p_m} = \widetilde{\chi}\widetilde{B} + i\widetilde{G}\widetilde{E}$, we can get

$$\tilde{\alpha} = \frac{\varepsilon_0 \overrightarrow{\alpha_1}}{\overleftrightarrow{I} + \varepsilon_0 c Z_0 k^4 \overrightarrow{\alpha_1} \overrightarrow{G_m}(r_m, r_e) \overrightarrow{u_2} \overrightarrow{G_m}(r_e, r_m)} \quad (S18)$$

$$\tilde{\chi} = \frac{\overrightarrow{u_2}/u_0}{\overleftrightarrow{I} + \varepsilon_0 c Z_0 k^4 \overrightarrow{u_2} \overrightarrow{G_m}(r_e, r_m) \overrightarrow{\alpha_1} \overrightarrow{G_m}(r_m, r_e)} \quad (S19)$$

$$\tilde{G} = -i \frac{\frac{\varepsilon_0 Z_0 k^2 \overrightarrow{\alpha_1} \overrightarrow{G_m}(r_m, r_e) \overrightarrow{u_2}}{u_0}}{\overleftrightarrow{I} + \varepsilon_0 c Z_0 k^4 \overrightarrow{\alpha_1} \overrightarrow{G_m}(r_m, r_e) \overrightarrow{u_2} \overrightarrow{G_m}(r_e, r_m)} = \left( -i \frac{\varepsilon_0 c k^2 \overrightarrow{u_2} \overrightarrow{G_m}(r_e, r_m) \overrightarrow{\alpha_1}}{\overleftrightarrow{I} + \varepsilon_0 c Z_0 k^4 \overrightarrow{u_2} \overrightarrow{G_m}(r_e, r_m) \overrightarrow{\alpha_1} \overrightarrow{G_m}(r_m, r_e)} \right)^* \quad (S20)$$

Consider the expression of $C = -\frac{\varepsilon_0 \omega}{2} Im(\boldsymbol{E}^* \cdot \boldsymbol{B})$, $G \to \frac{G}{\varepsilon_0}$ .

**Methods**

**FEM simulation:** All full wave numerical simulation were done by using finite element method (FEM, commercial software package, Comsol Multiphysics 5.0). The three dimensional (3D) chiral nanostructrue were put in a homogeneous surrounding medium or on substrate. Non-uniform meshes were used for formatting the object. The largest mesh was set less than $\lambda/6$. Perfect matched layer (PML) was used to minimize the scattering from the outer boundary. The incident light was set to 1V/m and propagates along the z axis. The total scattering cross sections were obtained by integrating the scattered power flux over an enclosed surface outside the 3D nanostructure, while the absorption cross sections were determined by integrating the Ohmic heating within the nanostructure. The circular dichroism of the system were calculated as the difference in extinction under left and right handed circularly polarized light ($CD = \sigma_L - \sigma_R$ ). The dipole power were from formula 16&17.

The extinction spectra, the dipole power spectra and CD spectra of the 3D chiral modes and corresponding electric dipole momentum Pe and magnetic dipole momentum Pm in real part were shown in Fig S1-S13. The nanostructure were shown in the inset of the extinction spectra, respectively. All the structure were 3D chiral plasmonic structure because we expected to investigate the applicability of the quantitative model. The results of the simulation have a few difference with the published paper may be because that the structure were not in full accord or the meshes in simulation had some difference.

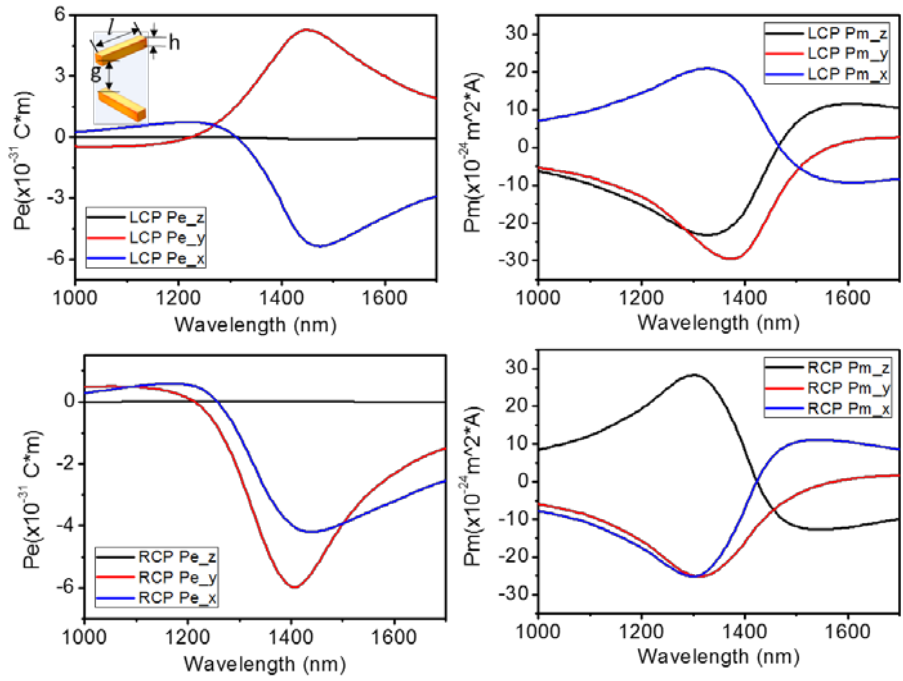

**Figure S1**. The electric dipole momentum **P$_e$** and magnetic dipole momentum **P$_m$** plotted in their x, y, z components with only real part for the structure in Figure 2. The figure is to show that the Pe and Pm have non zero components when they projects to each other, so the dot product is non zero.

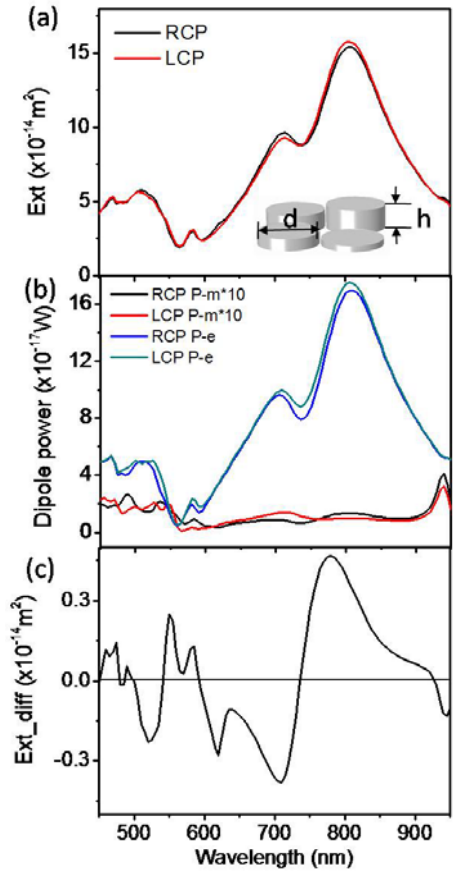

**Figure S2**. The extinction spectra (a), the dipole power spectra (b) and the extinction different (CD) spectra (c) of the Ag quasi-three-dimensional oligomers. (diameter: d = 100 nm, h = 10 nm, 20 nm, 30 nm, 40 nm, the gap between the oligomers: 2 nm)[3].

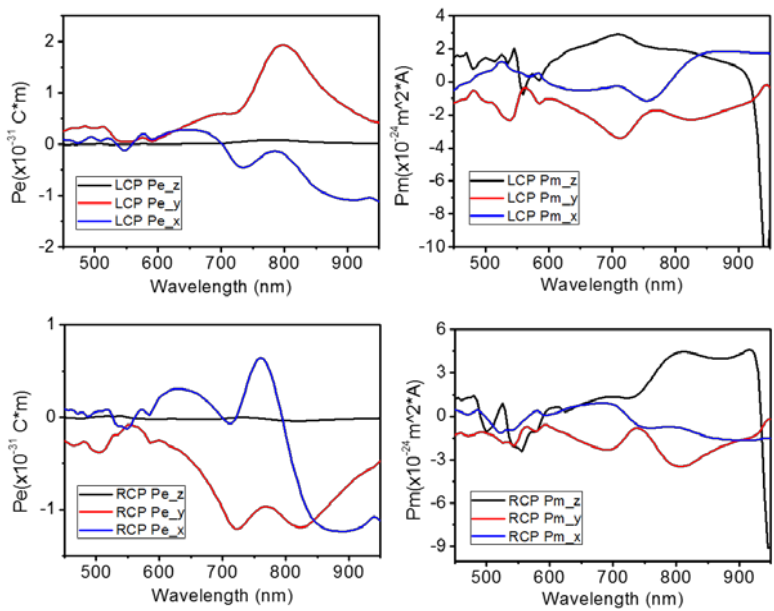

**Figure S3**. The electric dipole momentum $P_e$ and magnetic dipole momentum $P_m$ plotted in their x, y, z components with only real part for the structure in S2.

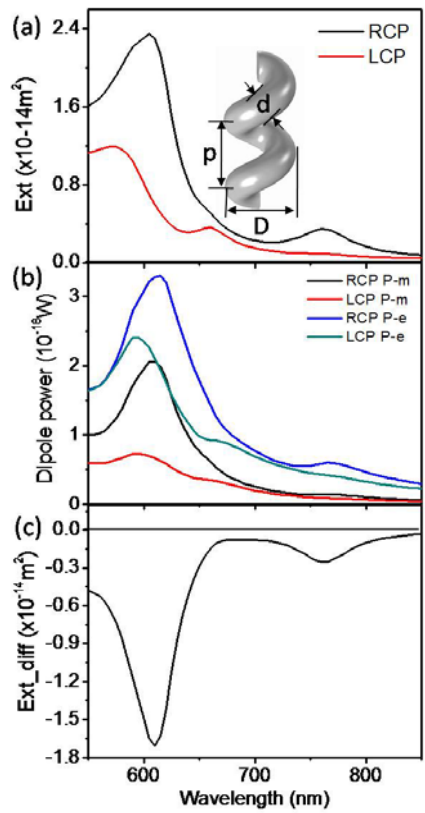

**Figure S4** The extinction spectra (a), the dipole power spectra (b) and the extinction different (CD) spectra (c) of the Cu plasmonic nanohelix. (major diameter: D = 36 nm, minor diameter: d = 28 nm, helical pitch: p = 60 nm)[4].

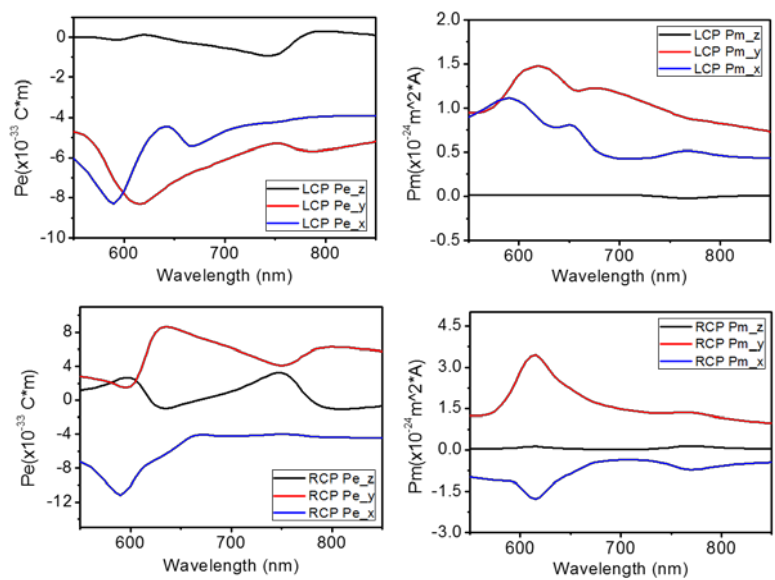

**Figure S5.** The electric dipole momentum $P_e$ and magnetic dipole momentum $P_m$ plotted in their x, y, z components with only real part for the structure in S4.

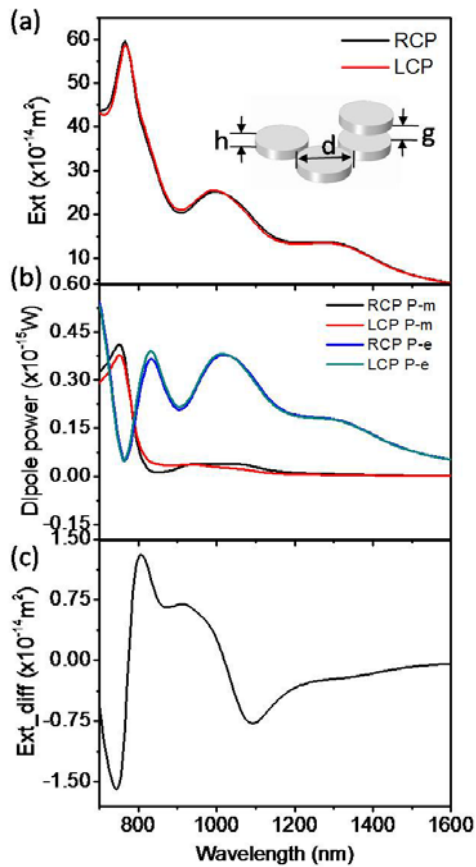

**Figure S6.** The extinction spectra (a), the dipole power spectra (b) and the extinction different (CD) spectra (c) of the Au 3D chiral plasmonic oligomers. (diameter: d = 100 nm, thickness of nanoparticle: h = 40 nm, the gap between the oligomers in borrom layer: s = 20 nm, the gap of two layers: g = 70 nm)[5].

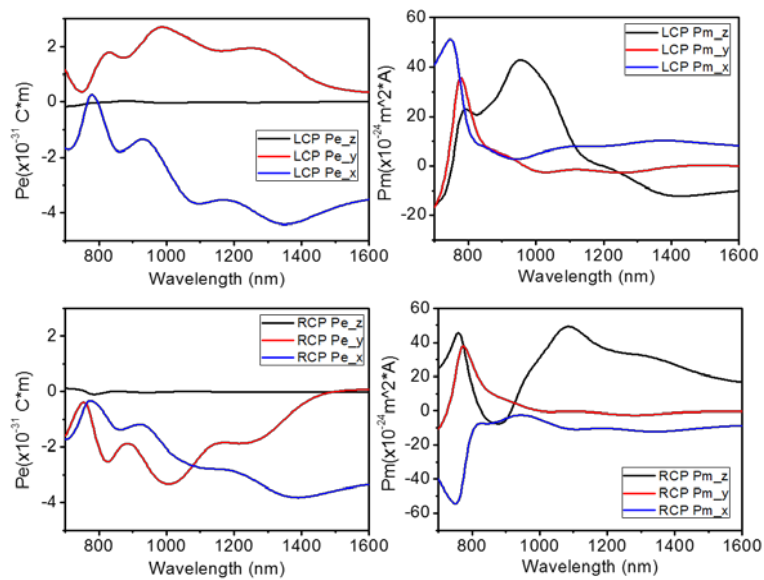

**Figure S7.** The electric dipole momentum $P_e$ and magnetic dipole momentum $P_m$ plotted in their x, y, z components with only real part for the structure in S6.

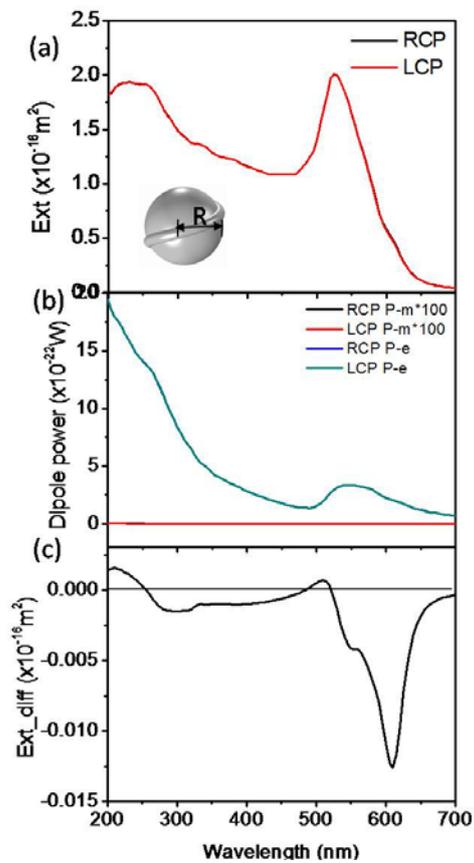

**Figure S8.** The extinction spectra (a), the dipole power spectra (b) and the extinction different (CD) spectra (c) of the Au chiral nanocrystals. (radius of sphere: R = 7 nm, radius of the twiser: 1 nm)[6]

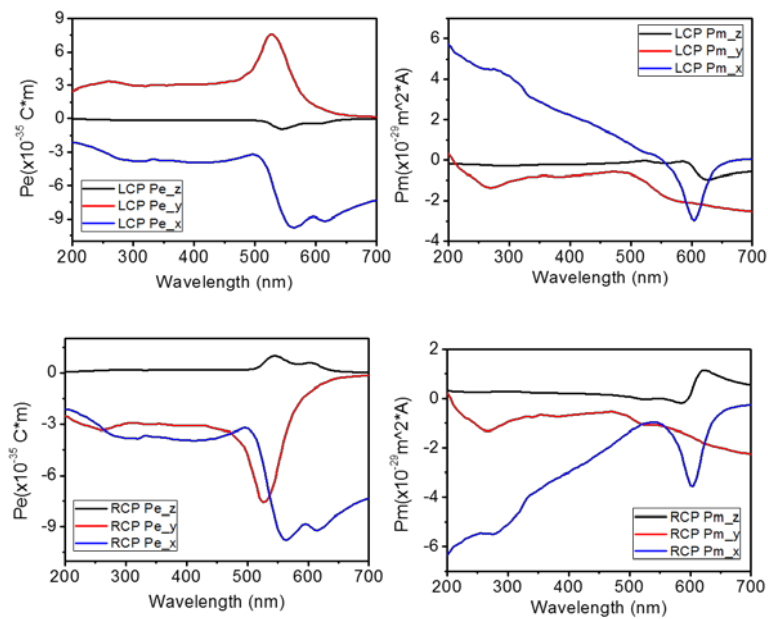

**Figure S9.** The electric dipole momentum $P_e$ and magnetic dipole momentum $P_m$ plotted in their x, y, z components with only real part for the structure in S8.

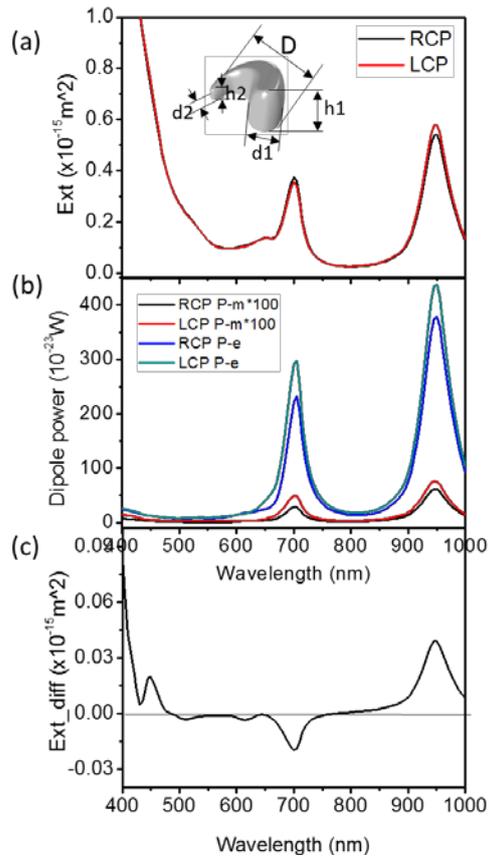

**Figure S10**. The extinction spectra (a), the dipole power spectra (b) and the extinction different (CD) spectra (c) of the spiral-type ramp nanostructures (outer diameter D = 22.5 nm, root diameter d1 = 11.3 nm, tip diameter d2 = 1.875 nm, the height at the root h1 = 17.5 nm, the height at the tip h2 = 2.5 nm).

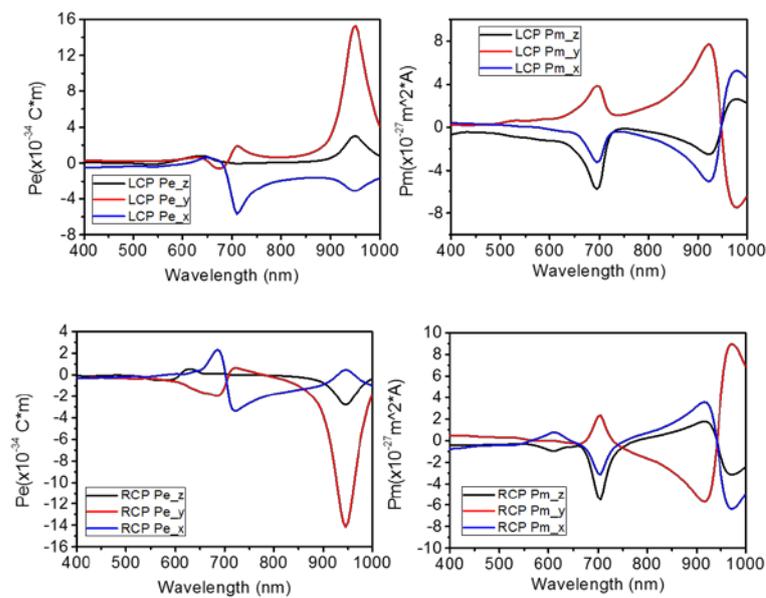

**Figure S11**. The electric dipole momentum $P_e$ and magnetic dipole momentum $P_m$ plotted in their x, y, z components with only real part for the structure in S10.

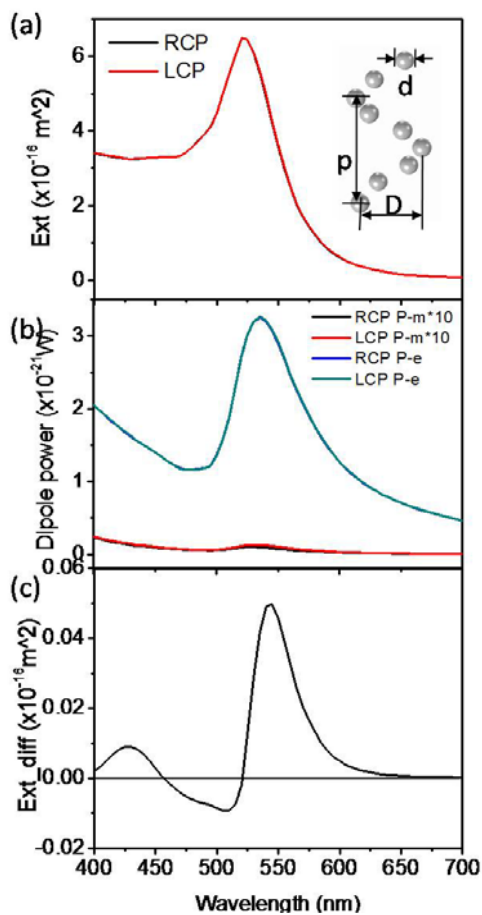

**Figure S12**. The extinction spectra (a), the dipole power spectra (b) and the extinction different (CD) spectra (c) of the Au nanoparticle helices. (nanosphere diameter: d = 10 nm, major diameter: D = 34 nm, helical pitch: p = 54nm)[7]

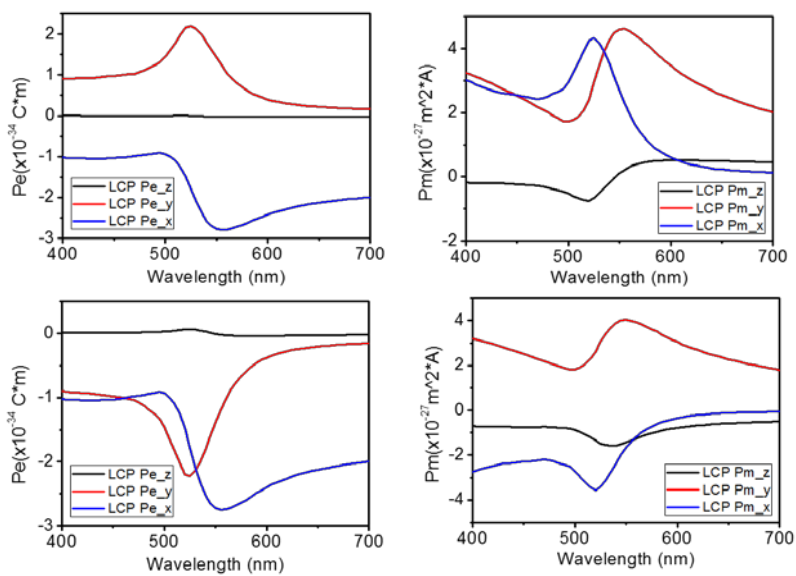

**Figure S13.** The electric dipole momentum $P_e$ and magnetic dipole momentum $P_m$ plotted in their x, y, z components with only real part for the structure in S12.